\documentclass[12pt,twoside,a4paper]{article}
\usepackage{epsfig}
\newcommand{\ddt}{\sigma^{diff}/\sigma_{\mathrm{tot}} }
\newcommand{\im}{\mathrm{Im\,}}
\begin{document}
\renewcommand{\baselinestretch}{0.9}
\title{
   \hfill {\bf\normalsize TAU 2605-99}\\
   {\bf Shadowing Corrections and Diffractive Production in DIS on
Nuclei}}
\author{{\bf
E.~Gotsman\thanks{e-mail:gotsman@post.tau.ac.il }~$\, ^a$,\quad
E.~Levin\thanks{e-mail: leving@post.tau.ac.il  }~$\, ^{a,b}$,\quad
M.~Lublinsky\thanks{e-mail: mal@techunix.technion.ac.il }~$\, ^c$,}\\
{\bf
U.~Maor\thanks{e-mail: maor@post.tau.ac.il }~$\, ^a$,\quad
K.~Tuchin\thanks{e-mail: tuchin@post.tau.ac.il}~$\, ^a$}\\[3mm]
{\it\normalsize $^a$HEP Department}\\
 {\it\normalsize School of Physics and Astronomy,}\\
 {\it\normalsize Raymond and Beverly Sackler Faculty of Exact Science,}\\
 {\it\normalsize Tel-Aviv University, Ramat Aviv, 69978, Israel}\\[2mm]
 {\it\normalsize $^b$Brookhaven National Laboratory}\\
 {\it\normalsize New York, USA}\\[2mm]
 {\it\normalsize $^c$Department of Physics}\\
 {\it\normalsize Technion -- Israel  Institute of Technology}\\
 {\it\normalsize Haifa 32000, Israel}}
 \date{}
\maketitle
\thispagestyle{empty}  
\vspace*{-1cm}
\begin{abstract}
We calculate, in the Glauber-Mueller approach,  the ratio of the 
 diffractive dissociation cross section  to the
total cross section on nuclei. We observe that shadowing corrections ( SC
) are significant, in the calculated   $\ddt$ ratio but they do not lead
to a smooth energy
dependence of the ratio  for a  proton target as was observed at HERA.
\end{abstract}

\vspace*{.5cm}
\newpage
\setcounter{page}{1}
\section{Introduction}

In this letter we discuss the energy dependence of 
  $\ddt$, where $\sigma^{diff}$ is the
cross section for the diffractive dissociation of a virtual photon
($\gamma^*$), and $\sigma_{tot}$ is the virtual photon nuclear 
total cross section. It  was observed at HERA 
that for DIS on a  nucleon  this ratio, when taken for
definite masses of diffractively produced hadrons, is almost
independent of $W$  (the c.m. energy of the virtual photon - proton ) , 
over a
wide range of  photon virtualities $Q^2$ \cite{HERADD}.  At present
there is no theoretical explanation of this intriguing phenomena, except
for
the quasi-classical  gluon field approach \cite{BUCH}. However, in this
model  
the total and diffractive cross sections
 do not depend on energy, in 
contradiction to the experimental data.
 Some attempts have been made to develop a successful  
phenomenology, which describes the experimental data fairly well
\cite{PH}.
 Recently, aiming to explain
this phenomenon {theoretically} , Yu.~Kovchegov and L.~McLerran \cite{KML}
have suggested
that shadowing corrections (SC)
 play a crucial role in the kinematical region of interest. They
derived  a formula for $\ddt$ which takes into account  the quark --
antiquark final state. In the following, we generalize
the   Kovchegov-McLerran
formula using the  Glauber-Mueller approach \cite{Glauber} to SC.
We take into  account  the  $q\bar{q}$ plus one extra gluon
final state without which we are  unable to explain the diffractive
production at HERA \cite{GLMSM}. In the same
spirit we  derive a formula for $\ddt$ for an arbitrary
nucleus.

For the case of a nucleus target, the Glauber-Mueller (G-M) approach can
be
justified theoretically \cite{braz2,KOV}, while for DIS on a
nucleon this approach is just one of the many formalisms used to estimate
shadowing corrections. The kinematic region where the G-M approach is
valid is well defined. In addition, as shown by Kovchegov \cite{KOV}, it
provides the initial conditions for the GLR evolution equation
\cite{GLR} in the full kinematic region of DIS.
 \footnote{
The theoretical proof
 assumed sufficiently heavy nuclei. Based on much 
 experience with ``soft" Pomeron models \cite{SM}, it appears 
 that the Glauber approach can be used for nuclei with A $\geq$ 30.}

In section~2 we calculate   $\ddt$ for a nucleus with $q\bar{q}$ and
$q\bar{q}G$ final states. In section~3 we give
numerical estimates for different nuclei and different
virtualities $Q^2$. In section~4 we conclude with a summary where we list  
some possible experimental consequences of our calculation.
An early rough calculation of diffraction off nuclei has been published 
in Ref.\cite{FS}.

\section{Shadowing Corrections}
It was shown in \cite{Glauber} that the total cross section for
the interaction of a virtual photon with the target can be
written as 
\begin{eqnarray}
\sigma_{tot}(\gamma^*+p)&=& \int dz\int
d^2r_{\bot}P^{\gamma^*}(z,r_\bot;Q^2)\sigma_{dipole}(x_B,
r_\bot)\\ &=& 2\int d^2b_\bot\int dz\int d^2r_\bot
P^{\gamma^*}(z,r_\bot;Q^2)\im a^{el}_{dipole}(x_B,r_\bot;b_\bot)
,\nonumber
\end{eqnarray}
where $P^{\gamma^*}(z,r_\bot;Q^2)$ denotes  the probability of finding  a
quark-antiquark pair of  size $r_\bot$ inside a virtual photon,
and  $\sigma_{dipole}$ is the cross section for the interaction of
a colour dipole with the target. The optical theorem has been used to
express
$\sigma_{tot}$ in terms of the forward elastic amplitude
$a^{el}_{dipole}$. The $s$-channel unitarity constraint reads 
\begin{equation}\label{unitarity}
2\im a^{el}(s,b_\bot)=|a^{el}(s,b_\bot)|^2+G^{in}(s,b_\bot) ,
\end{equation}
where $G^{in}$ corresponds to the contribution of all inelastic channels.
The total, elastic and inelastic cross sections of the  process
are given by
\begin{eqnarray}
\sigma_{tot} &=& 2\int d^2 b_\bot \im a^{el}(s,b_\bot)\label{un1}\\
\sigma_{el} &=& \int d^2 b_\bot | a^{el}(s,b_\bot)|^2\label{un2}\\
\sigma_{in} &=& \int d^2 b_\bot G^{in}(s,b_\bot)\label{un3} .
\end{eqnarray}
Using these relations,  the cross section for   diffractive
production of a $q\bar q$ pair  is 
\begin{equation}
\sigma^{diff}(\gamma^*p\rightarrow q\bar q)=\int d^2b_\bot
 P^{\gamma^*}(z,r_\bot;Q^2)|a^{el}_{dipole}(x_B,r_\bot;b_\bot)|^2 .
\end{equation}

To  obtain an estimate  for the amplitudes $a^{el}$
and $G^{in}$ we assume that the dipole-target amplitude is purely
imaginary, this  is a reasonable  approximation at high energies. Thus, 
the
general solution of the unitarity constraint, given by  
Eq.(\ref{unitarity}) can be
written
\begin{eqnarray}
a^{el}(x,r_\bot;b_\bot) &=& i\left(
1-e^{-\frac{\Omega^P(x,r_\bot;b_\bot)}{2}}\right)\label{ael}\label{GMF}\\
G^{in}_{dipole}(x,r_\bot;b_\bot) &=&
1-e^{-\Omega^P(x,r_\bot;b_\bot)} .
\end{eqnarray}

In the spirit of the Glauber-Mueller \cite{Glauber} approach, one
 may consider the function $\Omega^P$ as an amplitude for the
exchange of a  hard Pomeron (gluon ladder), which is given by
( see Fig.~1)
\begin{equation}
\Omega^P(x,r_\bot;b_\bot)=a^{el}_{dipole}(x,r_\bot;b_\bot)=
\frac{\pi^2\alpha_SA}{3}r_\bot^2
x_PG^{DGLAP}(x_P,\frac{4}{r_\bot^2})S(b_\bot) , 
\end{equation}
where $S(b_\bot)$ is the nucleus profile function for  which we take 
a Gaussian form
\begin{equation}
S(b_\bot)=\frac{1}{\pi R_A^2}e^{-\frac{b_\bot^2}{R_A^2}}.
\end{equation}
In the Woods-Saxon
parametrization \cite{WS}, the size of the nucleus is given by
$ R_A(WS)= 1.3\,
\mathrm{fm}\; A^{1/3}$, while in the  Gaussian parameterization $ R^2_A =
2/5
R^2_A(WS)$. 
 A full explanation of  all factors in Eq.(10) are given in Refs.
 \cite{braz2} and \cite{GLMPSI}.

In the Glauber-Mueller
approach at large $Q^2$, a useful    approximation  is to
replace the integration over z as follows \cite{Glauber,braz}
\begin{equation}
\int_0^1dz P^{\gamma^*}(z,r_\bot;q^2)=
\frac{4\alpha_{em}N_c}{3\pi^2Q^2}\frac{1}{r^4_\bot}\,\, ,
\end{equation}
which reproduces the  Kovchegov-McLerran formula\cite{KML,Nucleon}
\begin{equation} \label{bezcor}
\frac{\sigma^{diff}}{\sigma_{\mathrm{tot}}}=\frac{\int
d^2b_\bot\int\frac{dr_\bot^2}{r_\bot^4}
\left(1-e^{-\frac{\Omega^P(x,r_\bot;b_\bot)}{2}}\right)^2} {2\int
d^2b_\bot\int\frac{dr_\bot^2}{r_\bot^4}
\left(1-e^{\frac{-\Omega^P(x,r_\bot;b_\bot)}{2}}\right)} \; .
\end{equation}

\begin{figure}[htbp]
\begin{center}
\begin{tabular}{c}
\epsfig{file=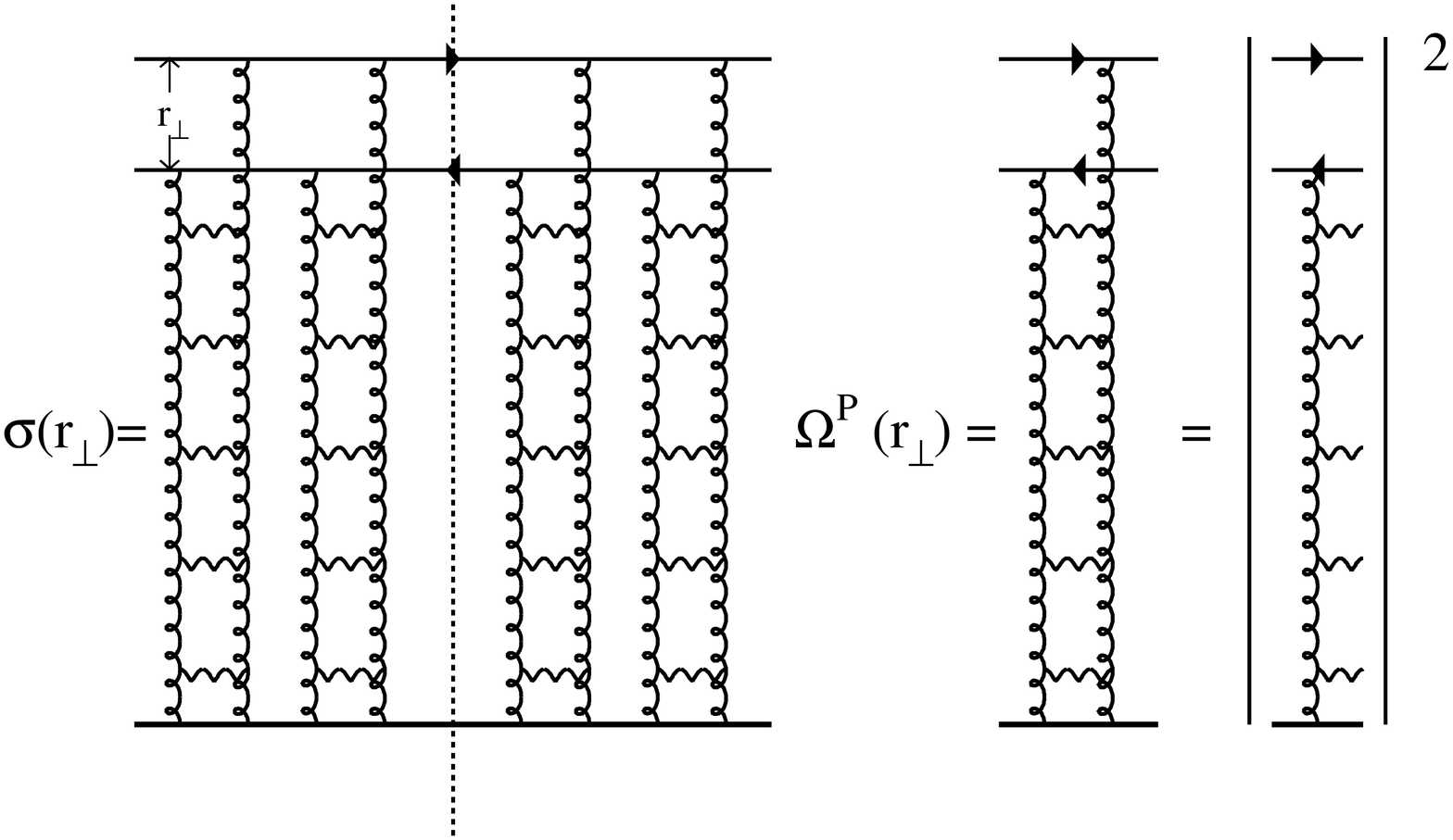,width=160mm,height=55mm}\\
Fig.1-a\\
\epsfig{file=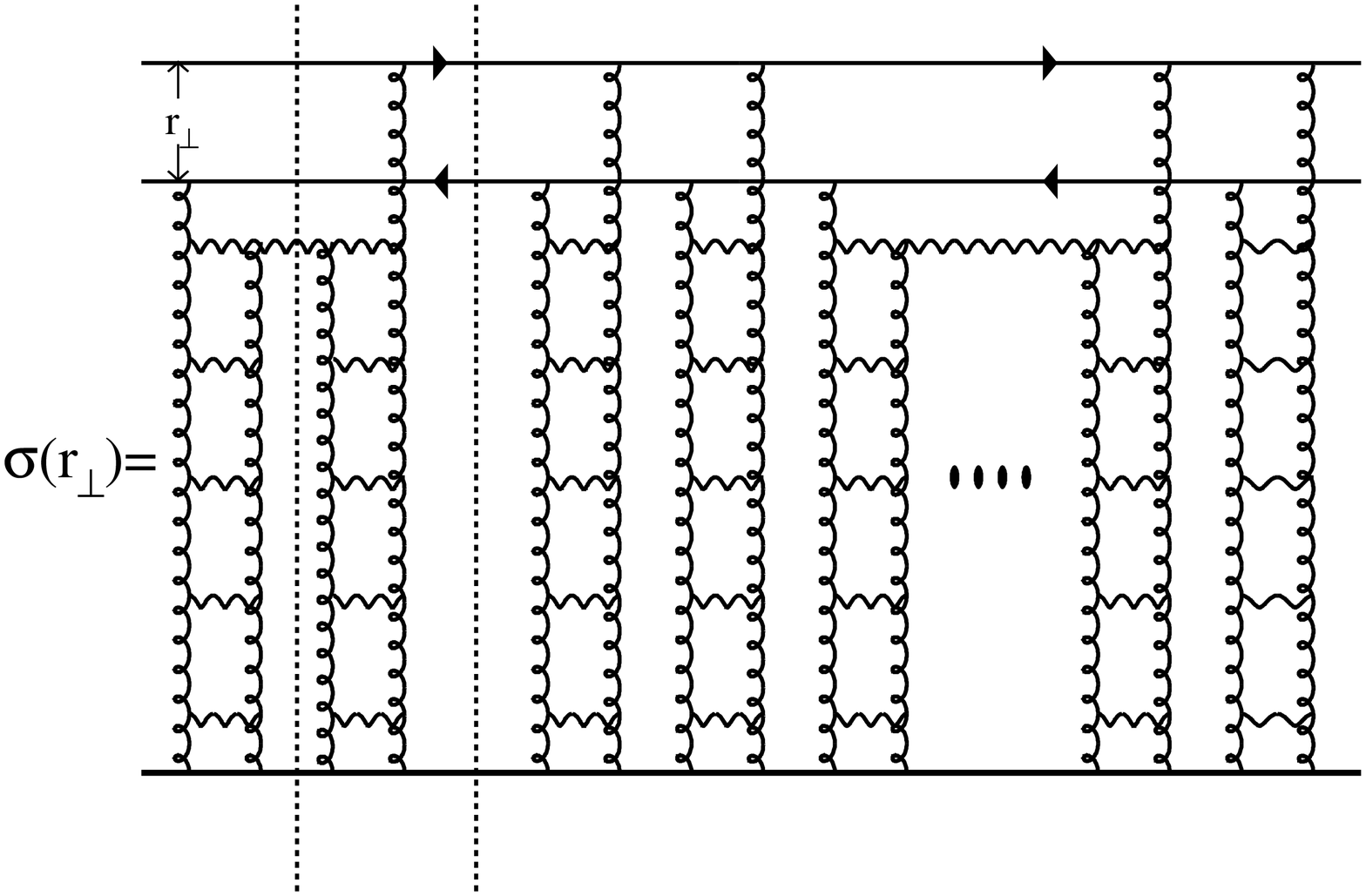,width=160mm,height=55mm}\\
 Fig.1-b
\end{tabular}
\end{center}
\caption{\it  
Total cross section  and diffractive cross sections for a  dipole -
nucleon
interaction in
the Glauber-Mueller approach (Fig. 1-a) and  in
the first interaction of the
G-M approach. The dashed lines show the diffraction
dissociation cuts. }
\end{figure}

There are two kinds of corrections to this simple
picture 
\cite{Nucleon} (see Fig.1): 
\newline 
 1)   As has been discussed \cite{Glauber},
 the G-M formula ( Eq.(~\ref{GMF}) describes the
rescattering of the fastest dipole interacting with  nucleus ( see Fig.1-a
). 
This rescattering depends on the opacity  $\Omega^P$, given by Eq.(9).
However, Eq.(10) depends on $ A x G^{DGLAP}$ which is also screened due to
the rescatterings of the fastest gluon in the gluon ladder (see Fig. 1-b
). To calculate this correction we iterate 
Eq.(~\ref{ael}), substituting instead of $\Omega^P$ the  expression from
the Glauber-Mueller formula \cite{Glauber, braz2}  for rescattering of
a gluon in a nucleus.
\begin{eqnarray}
&\Omega^P\,\,= \,\,
 \frac{1}{3}\pi^2\alpha_S\,A\,r_\bot^2 xG_N^{DGLAP}(x,\frac{4}{r_\bot^2})
S(b_\bot)&\\
&\rightarrow \Omega^{GM}=\frac{1}{3}\,\pi^2\alpha_S\,r_\bot^2\,
xG^{GM}(x,\frac{4}{r_\bot^2};b_\bot)\quad ,& \nonumber 
\end{eqnarray}
where
\begin{equation}
xG^{GM}(x,\frac{4}{r_\bot^2};b_\bot)=\frac{4}{\pi^2}\int_x^1\frac{dx'}{x'}
\int_{r_\bot^2}^\infty\frac{dr_\bot^{'2}}{r^{'4}_\bot}2
\left(1-e^{-\frac{\Omega_G(x',r_\bot';b_\bot)}{2}}\right).
\end{equation}
Note that the gluon ladder gets an additional colour factor
$\Omega_G=\frac{9}{4}\Omega^P$.
\newline
So far we have assumed  that the  diffraction dissociation final state
is merely a quark-antiquark pair plus the original nucleus ( $q\bar
q+A$ state ).

2)  The second improvement is  a correction
to this final state. For example, this  can be an  excited nucleus or
the  emission of an  extra gluon  (see the cuts in Fig.1-b ).  In
Ref.~\cite{Nucleon} we incorporate
these  correction
 using the AGK cutting
rules \cite{AGK}. The inelastic  cross section
of a dipole scattering off the target with the emission of  one extra
gluon in
the final state is given by
\begin{equation}
\sigma_{dipole}^{diff}(q\bar q\rightarrow q\bar qG)=\int d^2b_\bot
e^{-\Omega^{GM}(x,r_\bot;b_\bot)}\sigma_{GM}^{diff}\quad ,
\end{equation}
where
\begin{equation}
\sigma_{GM}^{diff}(b_\bot)=\frac{N_c\alpha_SA}{\pi}\int_x^1\frac{dx'}{x'}
\int\frac{dr_\bot^{'2}}{r_\bot^{'4}}
\left(1-e^{\frac{-\Omega_G(x',r_\bot';b_\bot)}{2}}\right)^2 ,
\end{equation}
provided that $\sigma_{MG}^{diff}\ll$$\Omega^{GM}\ll 1$.

  Introducing the shadowing corrections described above into
Eq.~(\ref{bezcor}), modifies  it as follows 
\begin{equation}\label{main}
R=\frac{\sigma^{diff}}{\sigma_{\mathrm{tot}}}\,\,= \,\,\frac{N}{D}\,\, ,
\end{equation}
where
\begin{eqnarray}
N &=& \int d^2b_\bot\int\frac{dr_\bot^2}{r_\bot^4}
\left\{\,\left( 1-e^{\frac{\Omega^{GM}(x,r_\bot;b_\bot)}{2}}\right)^2
\label{M1}\right. \\
 &+ &\left. e^{-\Omega^{GM}(x,r_\bot;b_\bot)}
\frac{2 \,r_\bot^2\alpha_S(\frac{1}{r_\bot^2}) }{3\pi}
\int_x^1\frac{dx'}{x'}\int_{r_\bot^2}^\infty\frac{dr^{'2}_\bot}{r_\bot^{'4}}
\left(1-e^{-\frac{\Omega_G(x',r_\bot';b_\bot)}{2}}\right)^2
\,\right\} \,\, \nonumber\\
D &=& 2\int d^2b_\bot\int\frac{dr_\bot^2}{r_\bot^4}
\left(1-e^{\frac{\Omega^{GM}(x,r_\bot;b_\bot)}{2}}\right) \,\,.\label{M2}
\end{eqnarray}
 We will use this formula in the numerical calculations (see
section~3). It should be stressed that  it is derived in the double log
approximation to the DGLAP
evolution equation. 
 It is instructive to compare this formula with 
successful phenomenology \cite{PH}, for example, with the Golec-Biernat and
Wusthoff approach where the  ideological basis is shadowing
corrections. The  Golec-Biernat and Wusthoff approach has no impact
parameter dependence and thus cannot be easily generalized for a
nucleus target.
  Deriving
  Eq.(20) from AGK cutting rules  leads to an
extra factor $exp( - \Omega^{GM} )$ in the second term of Eq.(20) which
does not appear in the  Golec-Biernat and Wusthoff approach. 

 For large nuclei   the  dependence on A of the
$q\bar qG$ term 
 is given by the factor $e^{-\Omega^{MG}(x,r_\bot;b_\bot)}$, 
which suppresses  it especially for large nuclei.  
\section{Numerical calculations}
We now  present the results of our numerical calculations
with Eq.~(\ref{main}). It is convenient  to carry out the integration
over $b_\bot$ analytically. We thus obtain
\begin{equation}\label{num1}
R=\frac{\sigma^{diff}}{\sigma_{\mathrm{tot}}}= \frac{
\int\frac{dr_\bot^2}{r_\bot^4}
\left[f^{q\bar q}(x,r_\bot)+
\frac{2\alpha_S(\frac{1}{r_\bot^2})\,r_\bot^2}{3\pi}
\int_x^1\frac{dx'}{x'}\int_{r_\bot^2}^\infty\frac{dr^{'2}_\bot}{r_\bot^{'4}}
f^{q\bar q G}(x,r_\bot;x',r_\bot')\right]}
{2\int\frac{dr_\bot^2}{r_\bot^4}f^{tot}(x,r_\bot)} ,
\end{equation}
where
\begin{eqnarray}
f^{q\bar q}(x,r_\bot)&=& 2E_1(\frac{1}{2}\Omega^{MG}(x,r_\bot))-
E_1(\Omega^{MG}(x,r_\bot))+C\nonumber\\
&+&\ln(\frac{1}{4}(\Omega^{MG}(x,r_\bot)))\label{num2}\\
f^{tot}(x,r_\bot;x',r_\bot')&=&
E_1(\frac{1}{2}\Omega^{MG}(x,r_\bot))+C+\ln(\frac{1}{2}\Omega^{MG}(x,r_\bot))
\label{num3}
\end{eqnarray}
\begin{eqnarray}
f^{q\bar q G}(x,r_\bot)&=&2E_1(\Omega^{MG}(x,r_\bot)+
\frac{1}{2}\Omega_G(x,r_\bot))-
E_1(\Omega^{MG}(x,r_\bot))\nonumber\\
&-&E_1(\Omega_G(x,r_\bot)+\Omega^{MG}(x,r_\bot))-
\ln(\Omega^{MG}(x,r_\bot))\nonumber\\
&+&
2\ln(\frac{1}{2}\Omega_G(x,r_\bot)+\Omega^{MG}(x,r_\bot))\nonumber
\\
&-&\ln(\Omega^{MG}(x,r_\bot)+\Omega_G(x,r_\bot))\,\, ,\label{num4}
\end{eqnarray}
where $C$ is the Euler constant, and $E_1$ is the exponential integral
function of the first kind.

Our goal  in this paper is to study shadowing corrections to DIS. It seems
reasonable to choose the  GRV94 parametrization \cite{GRV94} as  input for
the DGLAP gluon
structure function $xG^{DGLAP}$. Indeed, GRV94 was derived 
 by fitting   experimental data at
not too small $x$, so we can hope that the  effects of SC which are large
at 
small $x$ are not hidden in the  specific choice of the  initial
conditions in the   parametrization (as  is the case in the more
advanced global fits to the evolution equation 
\cite{GRV98} \cite{MRS99} \cite{CTEQ}).
Also, the  GRV94 parameterization, with its low initial virtuality
$Q_0^2$,
is  close to
the solution of DGLAP in the Leading Log approximation.
This is compatible with our master equation~(\ref{main}) which 
has been  written
in the same approximation.

To avoid double counting we do not utilize more modern parameterizations of
the structure functions which were determined using data at much lower
values of $x$ and $Q^{2}$, and hence are more likely to include some of
the screening corrections.

 Before discussing the results of our numerical calculation we 
would like to caution the reader that,
 we calculate the diffractive dissociation cross section
integrated over all produced masses, while in the experimental
data \cite{HERADD}  a mass cutoff
has been introduced.
 Calculations, using Eqs.~(\ref{num1}--\ref{num4}), are plotted
in Figs.~2, 3, 4, 5 and 6. Figs.~2 and 4 show the dependence of $R$
(defined
in 
Eq.~(21))  on the virtuality $Q^2$. 
We see that the  dependence of $R$ on $Q^2$ becomes steeper as $x$
decreases
from $x=10^{-3}$ to $x=10^{-5}$
over a wide range of $0.5\le Q^2\le 50$ GeV$^2$. This implies that we are
entering the saturation region, although the values that $R$ obtains even
for small $Q^2$ are far from the  unitarity limit $1/2$ (see
Eqs.(~\ref{un1}\,-\,\ref{un3})). 
Obviously, the SC increase with the number of
nucleons ($A$), since the parton density scales roughly as
$A^{1/3}$. Figs.~3 and 5 demonstrate the increase of the SC with $A$.

The same physical behavior is seen in Fig.~6. We see that the  energy
dependence of R is far from being flat, contrary to some early
expectations\cite{PH}, which were based on the fact that experimentally
measured R for a single nucleon is flat for mass bins up to
15~GeV\cite{HERADD}. The main difference of diffractive
production in DIS on a nucleon target  compared to that on a nucleus 
target
 is the fact that the first depends crucially on the experimental 
cutoff of the produced mass.
At $Q^2\sim 1$ GeV$^2$ R reaches a value of 0.3 at $W=220$
GeV for DIS on a nucleus as heavy as $A=200$.  Note, that at
values of $Q^2>10$ GeV$^2$ shadowing corrections are  hardly seen in the  
HERA kinematical region.  

Finally we draw attention to the Fig.~7 where contributions of $q\bar q$
and $q\bar q G$ states to $R$ are shown separately. As seen, the
contribution of
$q\bar q G$ is almost independent of energy, while $q\bar q$ is the main
source of energy dependence of $R$.

The fact that $\ddt$
is rather small (significantly less than 1/2) for all nuclei  and it
varies substantially  with
energy,  gives rise to a problem of describing HERA data within
the  Kovchegov-McLerran formula.   We discuss this problem in another 
publication \cite{Nucleon}. 

We would like to stress that our model  completely ignores all corrections
to the final state (apart from the radiation of gluons). It may
turn out that the contribution of some of these, e.g. excitation of
the nucleus, is rather large and  can change our results. However it seems
 unlikely that our conclusions would be changed significantly. 

\section{Conclusions}
In this letter we studied the influence of shadowing corrections in DIS on
nuclei. To achieve  this goal  the generalization of the 
Kovchegov-McLerran formula 
(\ref{main}) was derived in the
Glauber-Mueller approach. This  means that we assume  that 
a diffractively produced $q\bar q$ state  
scatters off the nucleons incoherently, and we only  keep the Leading Log
terms in the formulae for the cross sections.

Our master equation (\ref{main}) makes it possible to calculate the ratio
of the
diffractive dissociation cross section of quark-antiquark pair to the
total cross section. We take into account  the possible contamination of
the
final state  caused by  the radiation of a single gluon. This
contribution is  almost independent of energy, and thus we expect that it
is $q\bar q$ contribution which saturates the ratio $R$.

The results of the calculation show that SC start to play a significant 
role in
nuclear DIS diffraction in the HERA kinematical region.
We thus expect that clear experimental signatures of  SC will  
 be found in
 DIS  experiments
with nuclei at HERA and BNL.

\section{Acknowledgements}
   Two of the authors E.L. and U.M.  thank the Nuclear Theory Group
at BNL for their hospitality, and  Yura Kovchegov  and  Larry McLerran
 for
helpful discussions.

   This research was supported in part by the Israel Science Foundation,
founded by the Israel Academy of Science and Humanities, and by the United
States-Israel BSF Grant \# 9800276.

%
%
\begin{figure}
\begin{tabular}{cc}
$\mathbf{A}${\bf =30}
&
$\mathbf{A}${\bf =100}\\[-5mm]
\epsfig{file=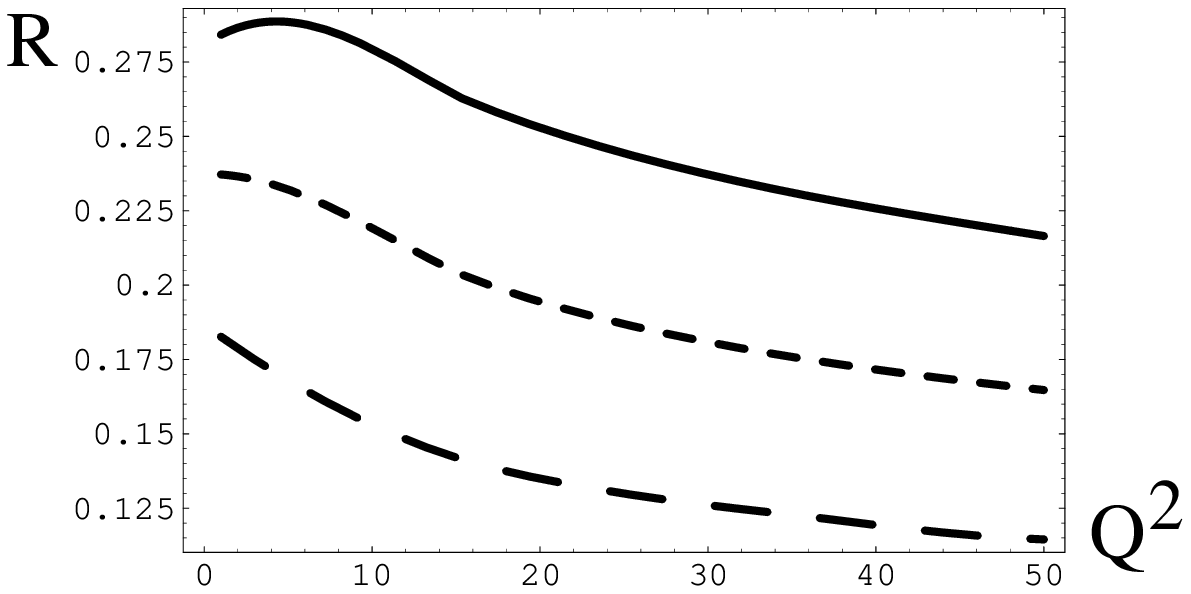,width=60mm,height=50mm}
&
\epsfig{file=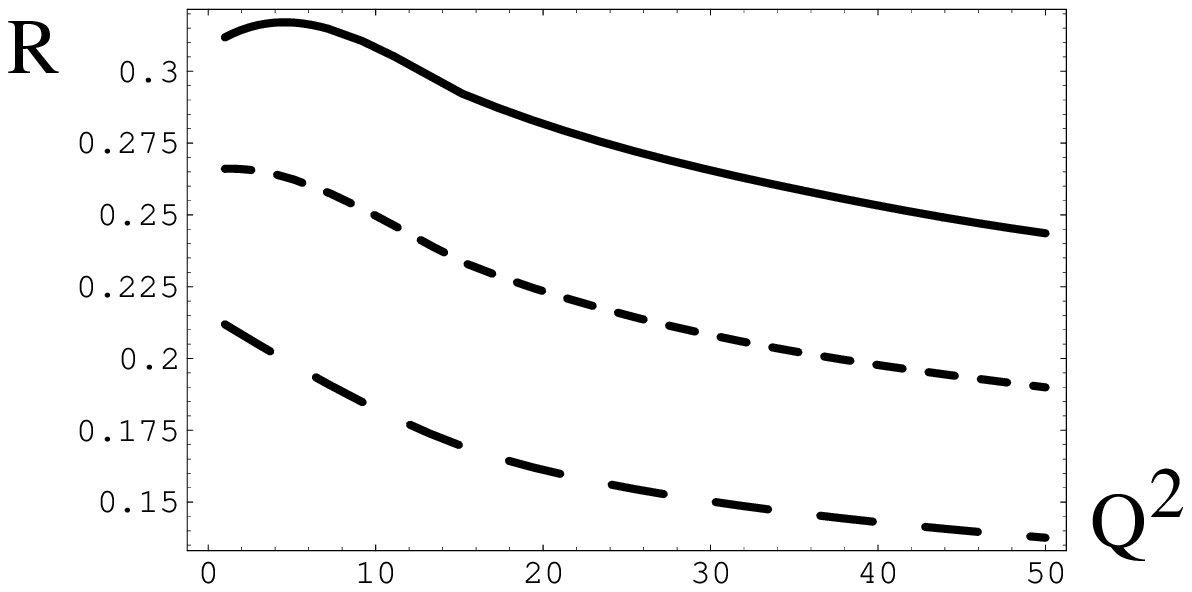,width=60mm,height=50mm}\\
$\mathbf{A}${\bf =200}
&
$\mathbf{A}${\bf =300}\\[-5mm]
\epsfig{file=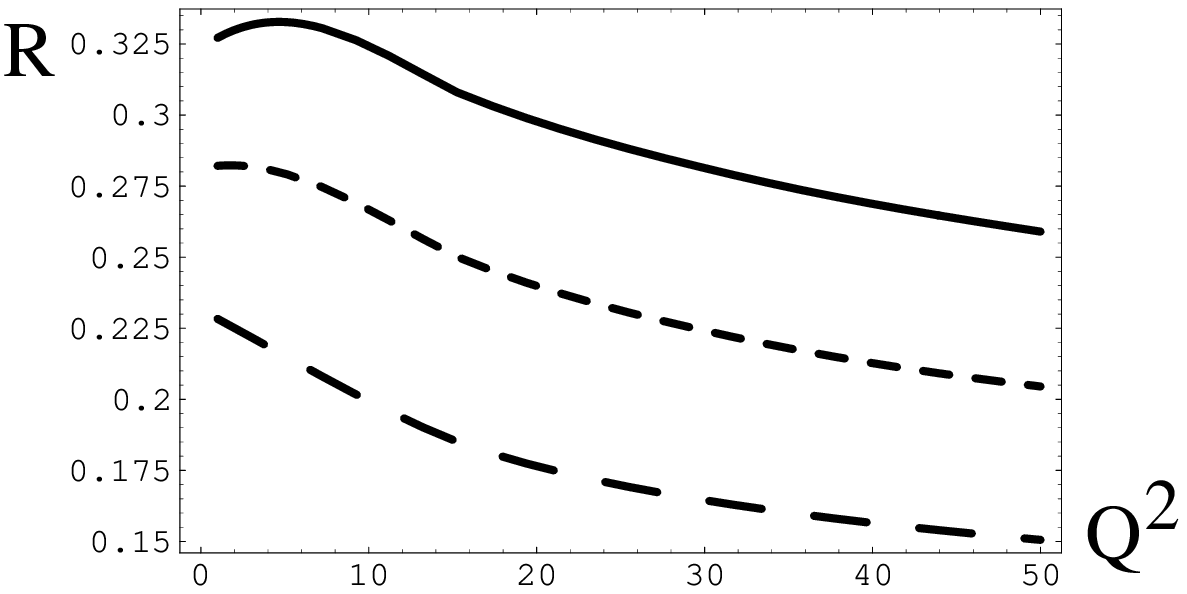,width=60mm,height=50mm}
&
\epsfig{file=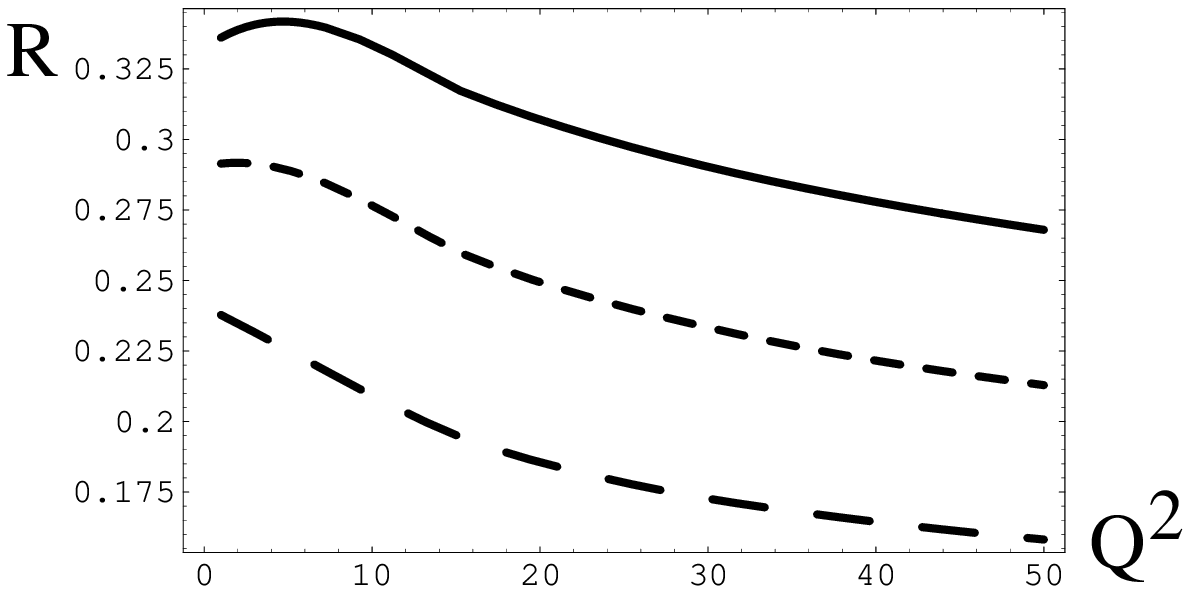,width=60mm,height=50mm}
\label{fig1}   
\end{tabular}
\vspace{5mm}
\caption{\footnotesize{\it R as a function of the
virtuality $Q^2$ for different $x$ and $A$. From the lower dashed curve
to the solid one: $x=10^{-3}$, $x=10^{-4}$, $x=10^{-5}$ .}}
\end{figure}
\begin{figure}
\begin{flushleft}
\begin{tabular}{cc}
$\mathbf{Q^2\,\,=\,\,1 \,GeV^2}$ & $\mathbf{Q^2\,\,=\,\,5
\,GeV^2}$\\[-5mm]
\epsfig{file=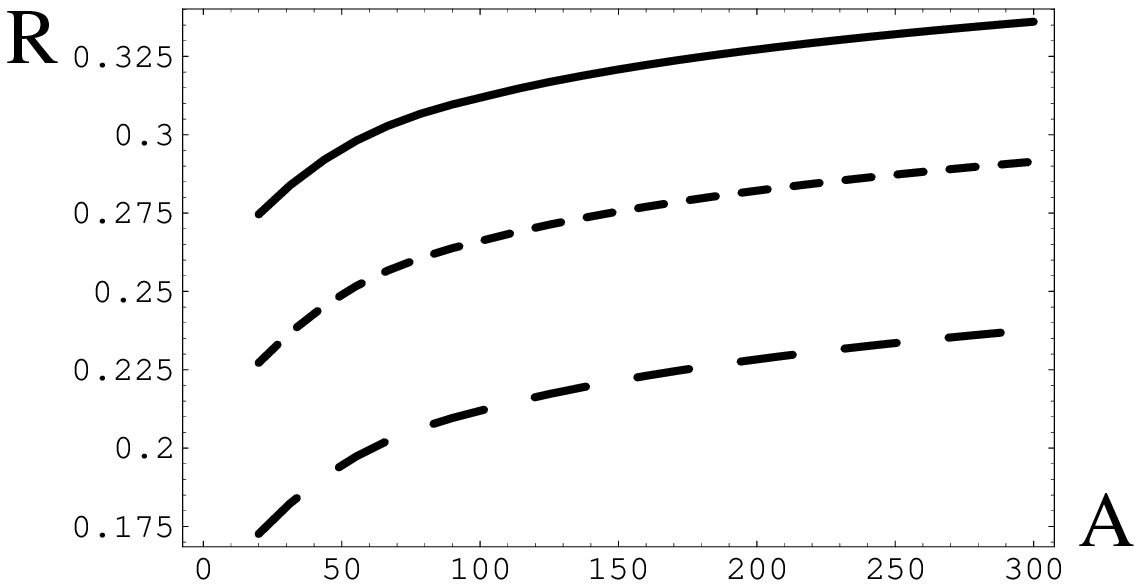,width=60mm,height=50mm}
&
\epsfig{file=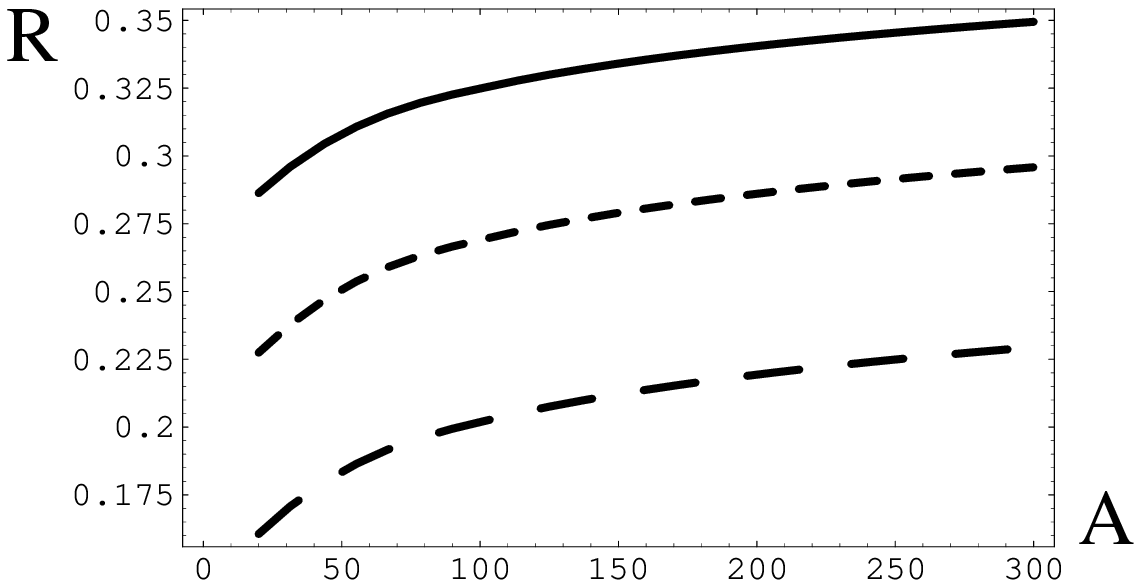,width=60mm,height=50mm}\\
$\mathbf{Q^2\,\,=\,\,12 \,GeV^2}$ & $\mathbf{Q^2\,\,=\,\,30 
\,GeV^2}$\\[-5mm]
\epsfig{file=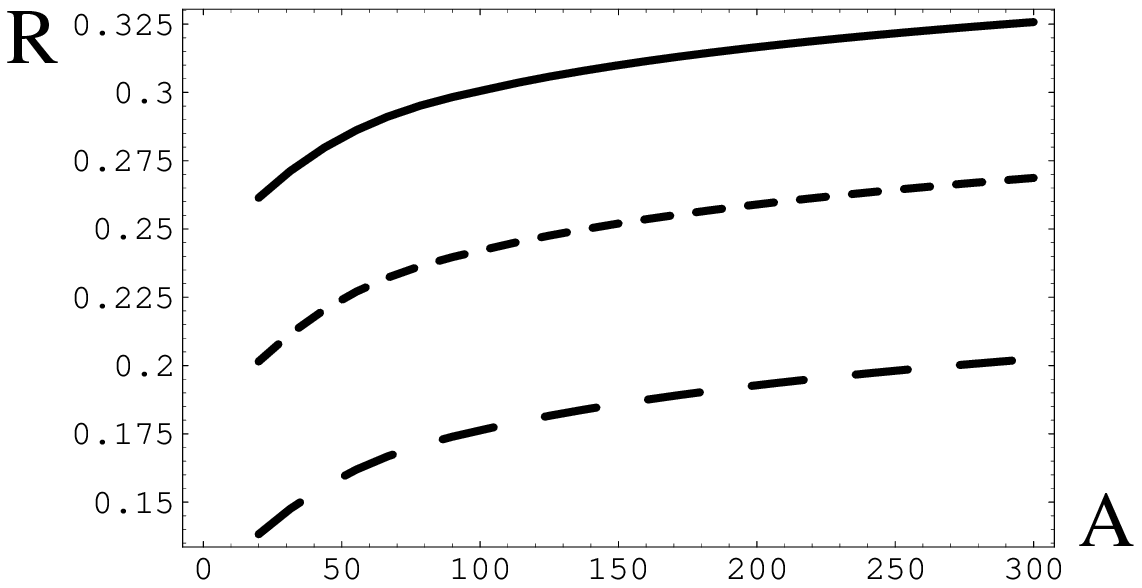,width=60mm,height=50mm}
&
\epsfig{file=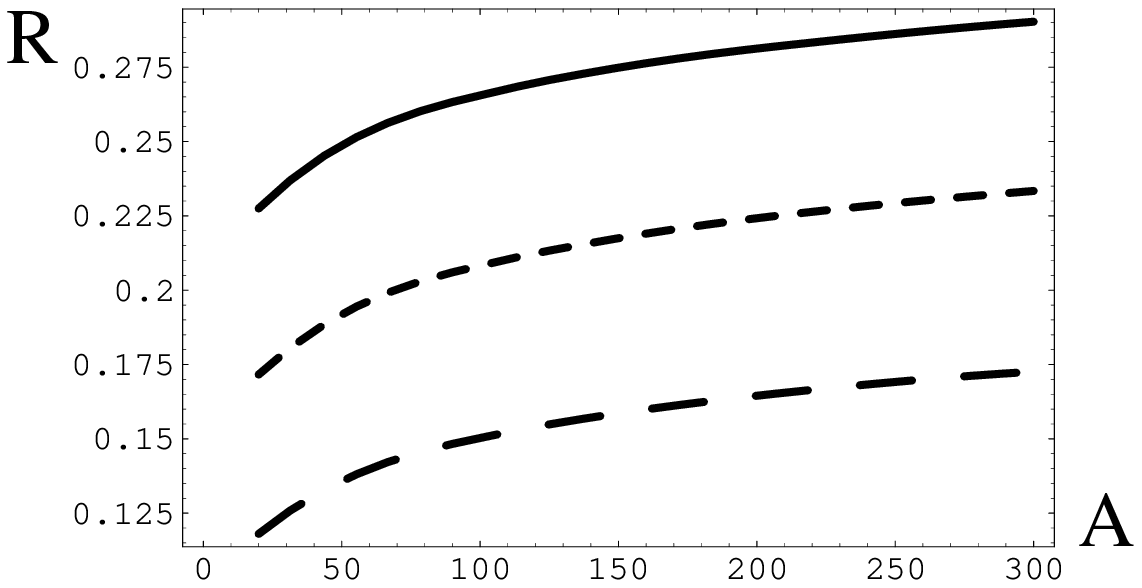,width=60mm,height=50mm}\\
$\mathbf{Q^2\,\,=\,\,50 \,GeV^2}$ & \\[-5mm]
\epsfig{file=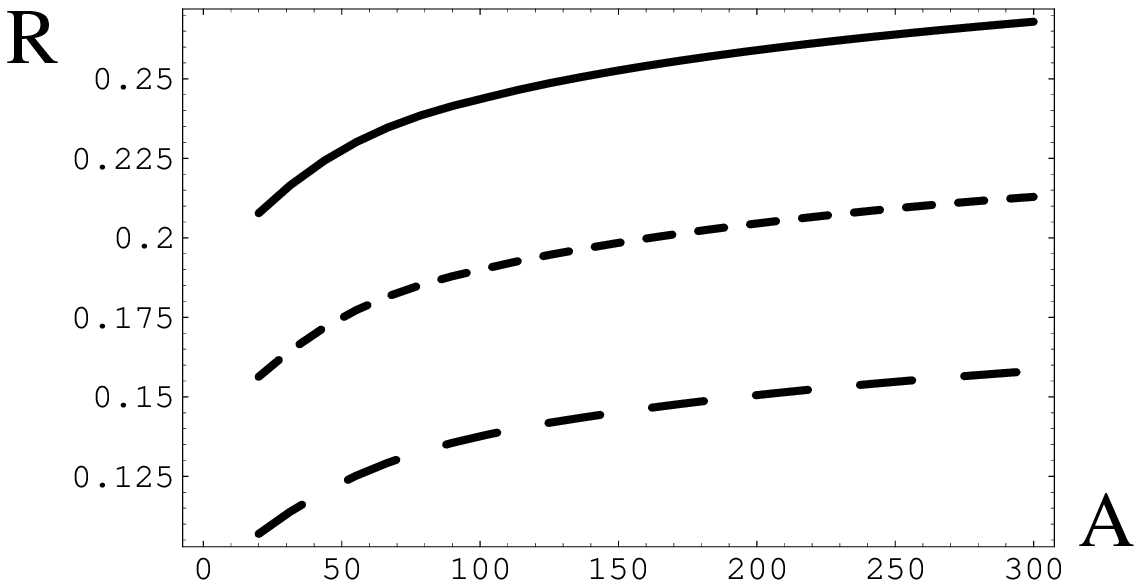,width=60mm,height=50mm}
&
\end{tabular}
\end{flushleft}
\vspace{0.17cm}
\caption{\footnotesize{\it R as a function of the 
number of nucleons $A$ for different $x$ and $Q^2$. From the lower dashed
curve to the solid one: $x=10^{-3}$, $x=10^{-4}$, $x=10^{-5}$..}}
\label{fig4}
\end{figure}

\begin{figure}
\begin{flushleft}
\begin{tabular}{cc}
\epsfig{file=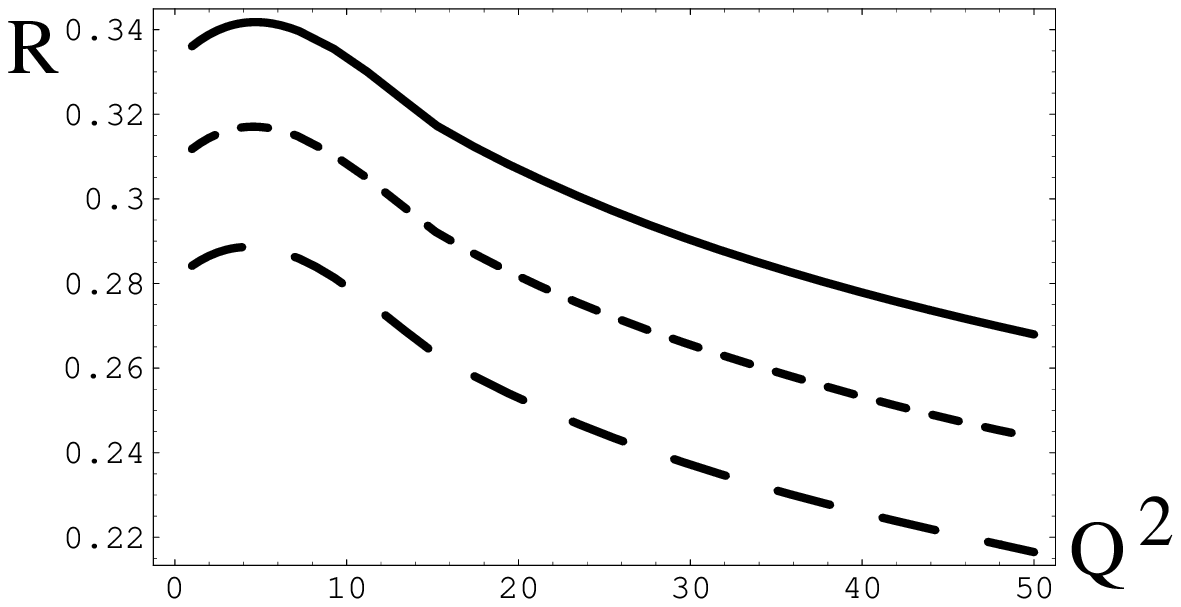,width=60mm,height=50mm}
&
\epsfig{file=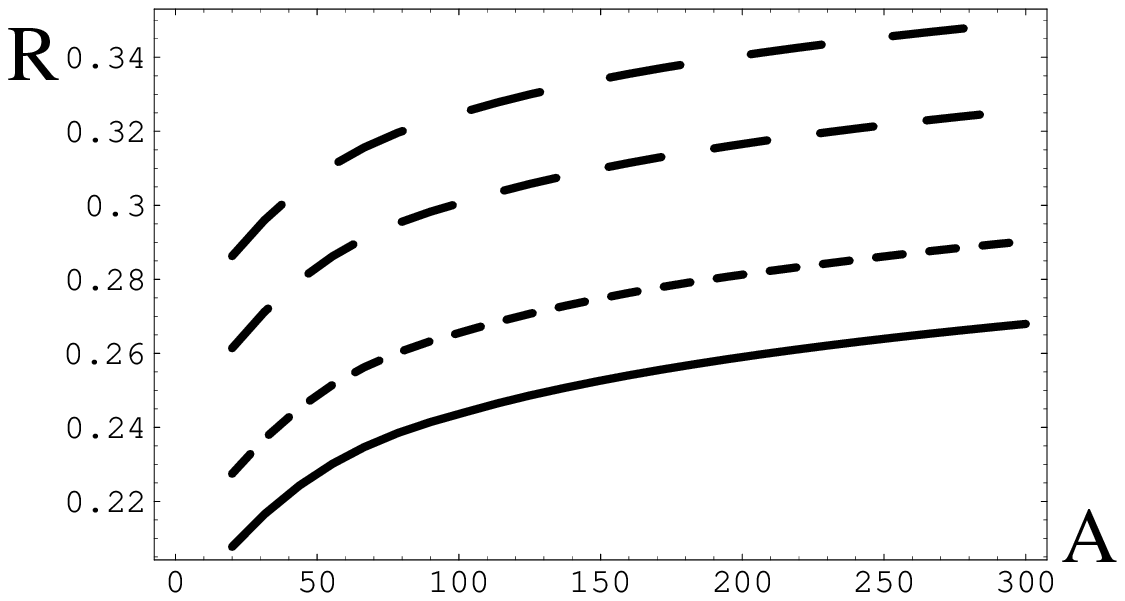,width=60mm,height=50mm}
\label{fig3}
\end{tabular}
\end{flushleft}
\vspace{0.17cm}
{\hspace*{0.1cm}{\footnotesize{ 
\begin{tabular}{cc}
Figure 4:{\it R as a function of the virtuality $Q^2$} &
Figure 5:{ \it  R as a function of number of}\\
{ \it at $x=10^{-5}$. From the lower dashed}&
{\it nucleons $A$ at $x=10^{-5}$. From the}\\
{\it curve to the solid one:}&
{\it upper dashed curve to the solid one:}\\
$A$\,=\,30,\,100,\,300. &
$Q^2$\,=\,1,\,5,\,12,\,30,\,60 GeV$^2$.
\end{tabular}
}}}
\end{figure}

\setcounter{figure}{5}

\begin{figure}
\begin{flushleft}
\begin{tabular}{cc}
$\mathbf{Q^2\,\,=\,\,1 \,GeV^2}$ & $\mathbf{Q^2\,\,=\,\,12
\,GeV^2}$\\[-5mm]
\epsfig{file=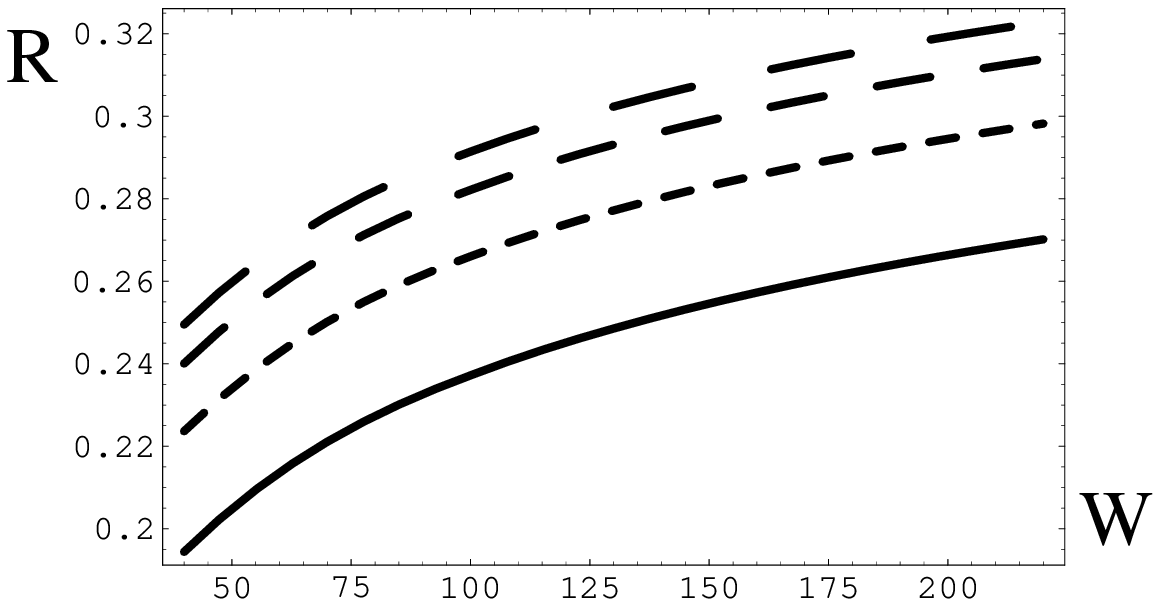,width=60mm,height=50mm}
&  
\epsfig{file=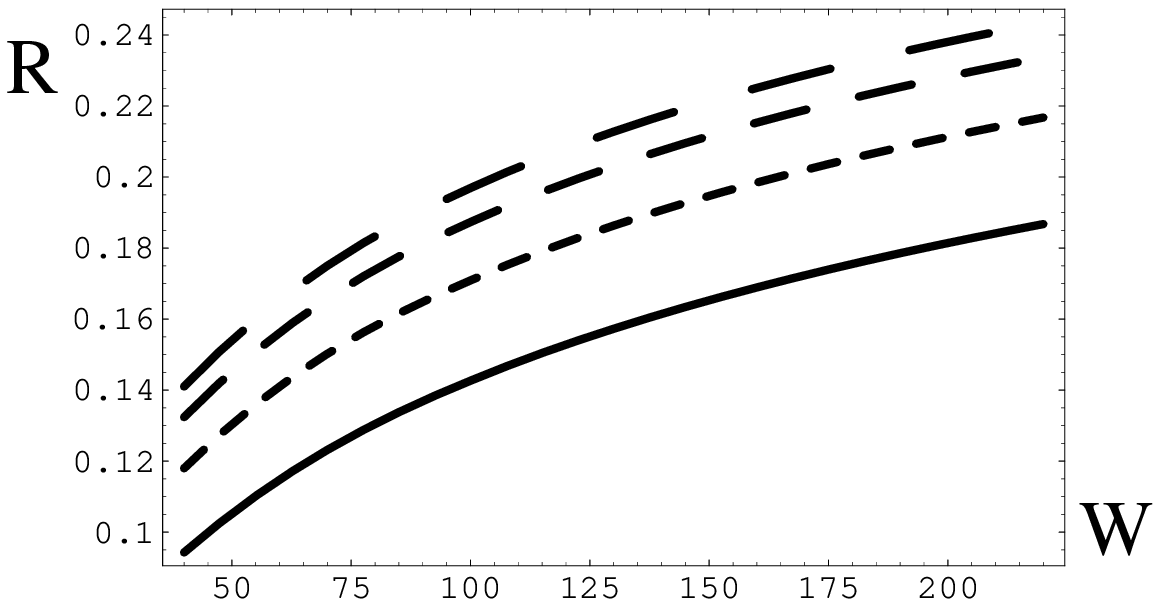,width=60mm,height=50mm}\\
$\mathbf{Q^2\,\,=\,\,30 \,GeV^2}$ & $\mathbf{Q^2\,\,=\,\,60
\,GeV^2}$\\[-5mm]
\epsfig{file=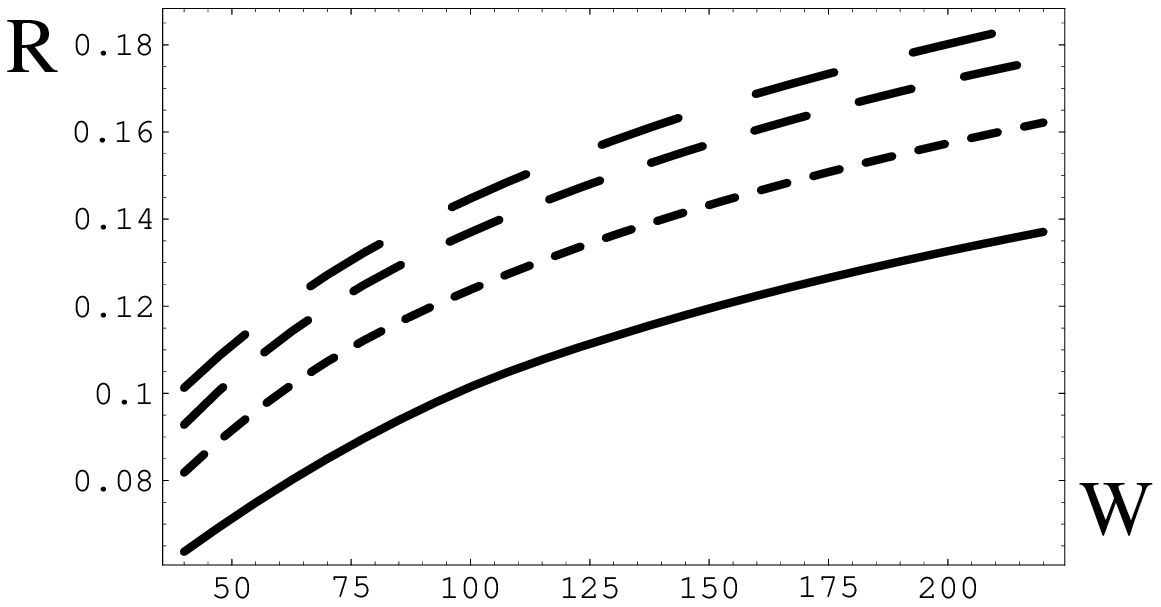,width=60mm,height=50mm}
&
\epsfig{file=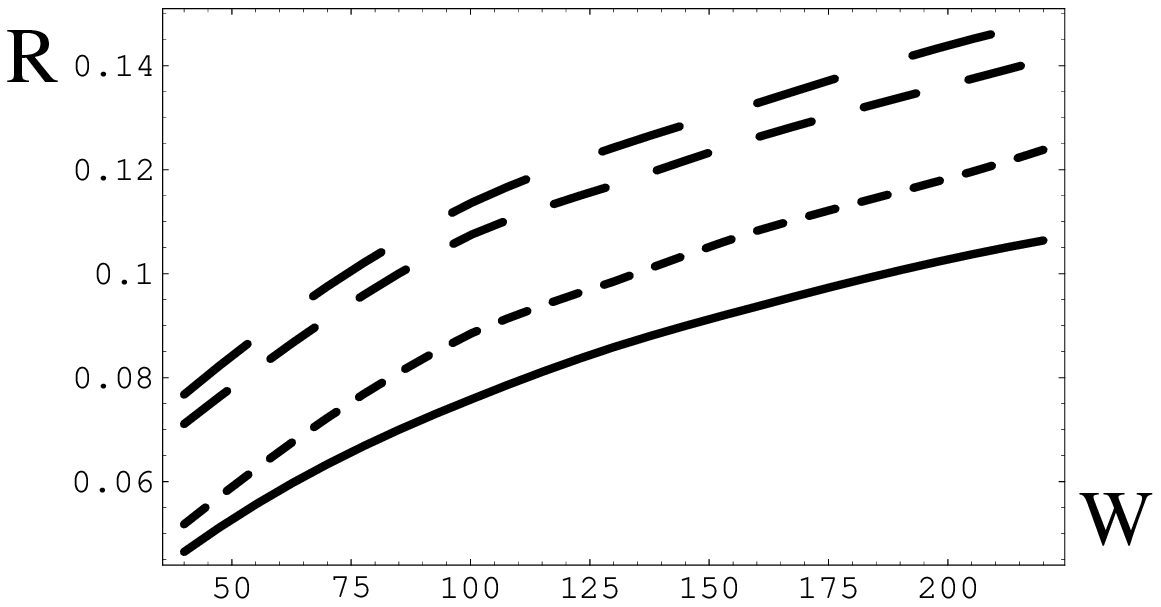,width=60mm,height=50mm}
\end{tabular}
\end{flushleft}
\vspace{0.17cm}
\caption{\footnotesize{\it R as a function of the
energy $W$ for different $A$ and $Q^2$. From the lower solid curve
to the upper dashed one: $A$\,=\,30,\,100,\,200,\,300.}}   
\label{fig5}
\end{figure}
\begin{figure}
\begin{flushleft}
\begin{tabular}{cc}
$\mathbf{A=30\quad Q^2=12 GeV^2}$& $\mathbf{A=200\quad
Q^2=12GeV^2}$\\[-5mm]
\epsfig{file=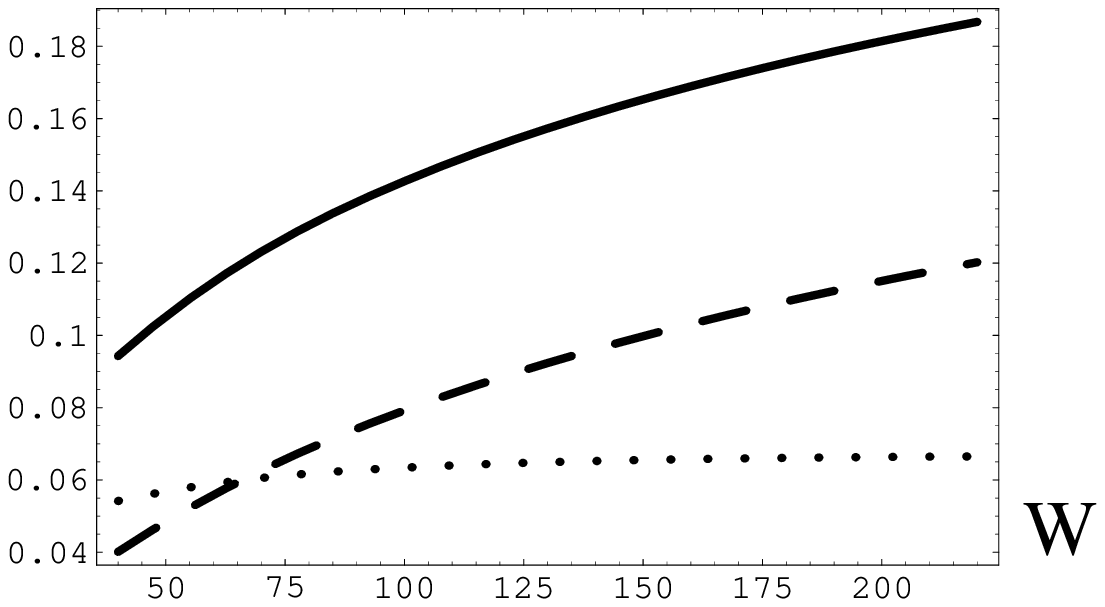,width=60mm,height=50mm}
&
\epsfig{file=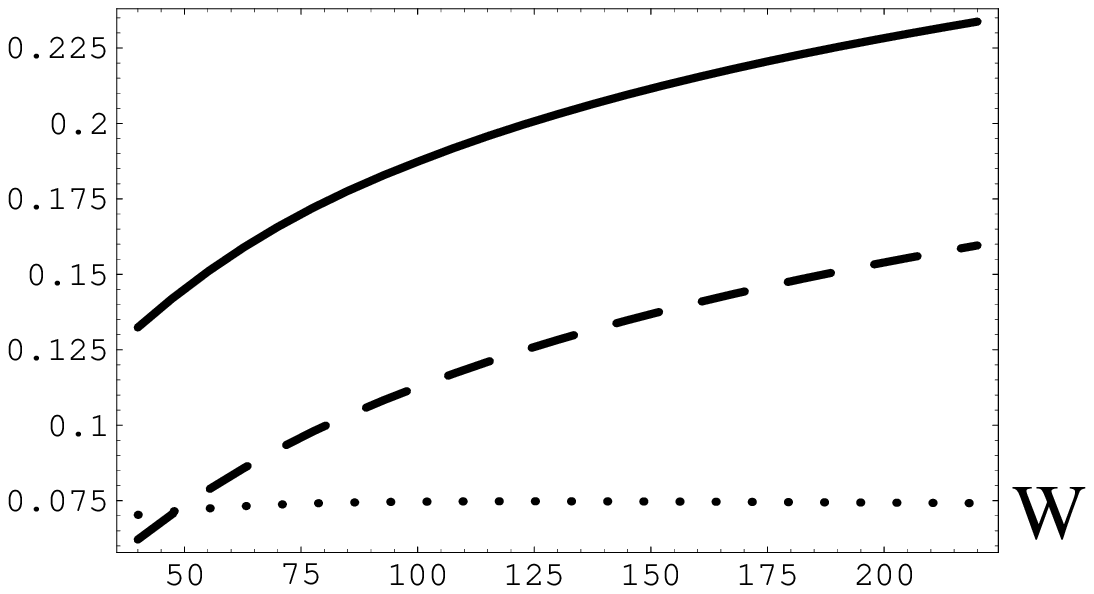,width=60mm,height=50mm}\\
$\mathbf{A=30\quad Q^2=60GeV^2}$ & $\mathbf{A=200\quad
Q^2=60GeV^2}$\\[-5mm]
\epsfig{file=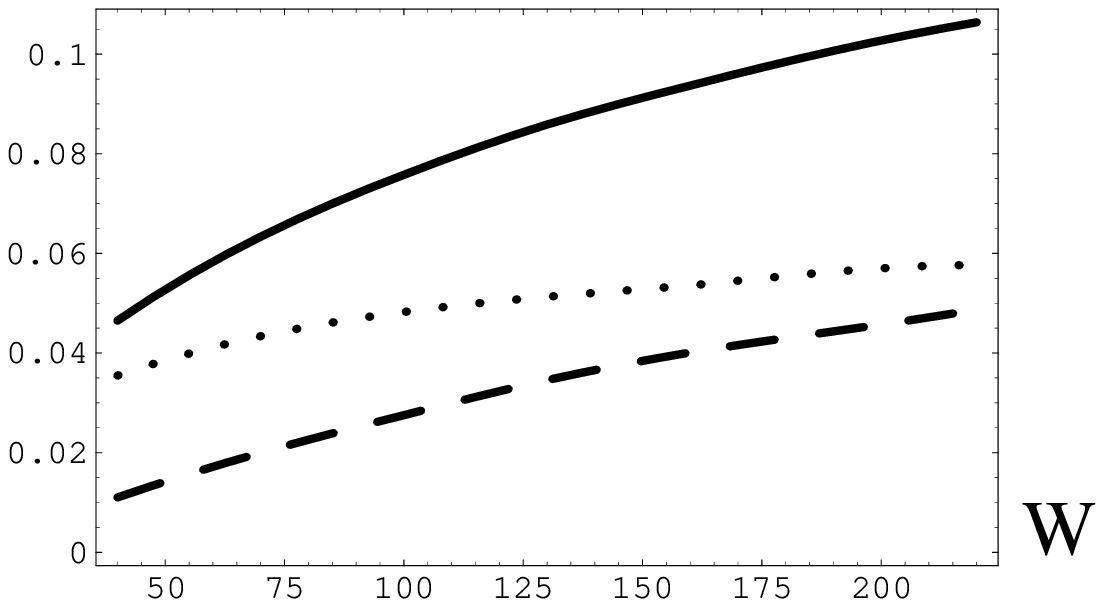,width=60mm,height=50mm}
&
\epsfig{file=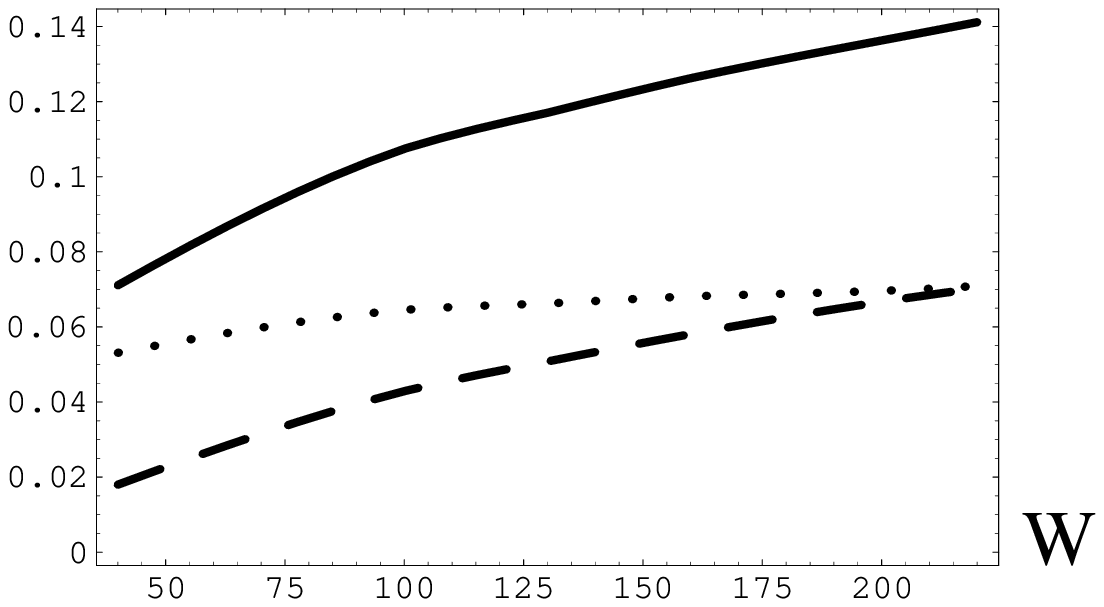,width=60mm,height=50mm}
\end{tabular} 
\end{flushleft}  
\vspace{0.17cm}
\caption{\footnotesize {\it Different contributions to R as a function of
the energy W for different $Q^2$ and A. Dashed line: contribution of the 
$q\bar q$ pair, dotted line: contribution of the $q\bar q G$ state, solid
line - total contribution.
}}
\label{fig6}
\end{figure}

%

\end{document}